\documentclass{article}
\usepackage[utf8]{inputenc}

\usepackage{amssymb}
\usepackage{amsmath}
\usepackage{amsthm}
\usepackage{graphicx}
\usepackage{booktabs}
\usepackage{siunitx}
\usepackage{hyperref}

\title{On the Difficulty of Evaluating Baselines \\
{\large A Study on Recommender Systems}}

\newcommand\rmseurbm{0.823}
\newcommand\rmsersvd{0.8256}
\newcommand\rmsebpmf{0.8197}
\newcommand\rmsebiasmf{0.803}
\newcommand\rmsesvdfeature{0.7907}
\newcommand\rmsealswr{0.7830}
\newcommand\rmseapg{0.8101}
\newcommand\rmsedfc{0.8067}
\newcommand\rmsegsmf{0.8012}
\newcommand\rmseiautorec{0.782}
\newcommand\rmsellorma{0.7815}
\newcommand\rmsewemarec{0.7769}
\newcommand\rmseicfnpp{0.7754}
\newcommand\rmsenade{0.771}
\newcommand\rmsempma{0.7712}
\newcommand\rmsesma{0.7682}
\newcommand\rmsegloma{0.7672} 
\newcommand\rmseermma{0.7670}
\newcommand\rmseadaerror{0.7644}
\newcommand\rmsemrma{0.7634}

\newcommand\rmsebfmui{0.7633}
\newcommand\rmsebfmuit{0.7587}
\newcommand\rmsebfmuiui{0.7563}
\newcommand\rmsebfmuitui{0.7523}
\newcommand\rmsebfmuituiiu{0.7485}
\newcommand\rmsefmui{0.7720}
\newcommand\rmsefmuimed{0.7756} 

\newcommand\rmseNFprobefunksvd{0.93}
\newcommand\rmseNFtestpatarek{0.9094}
\newcommand\rmseNFprobePT{0.9156}
\newcommand\rmseNFtestPT{0.9088}
\newcommand\rmseNFtestbgSGD{0.9028}
\newcommand\rmseNFtestZhou{0.8985}
\newcommand\rmseNFprobeGravity{0.9190}
\newcommand\rmseNFtestPilaszy{0.9018}
\newcommand\rmseNFtestYehuda{0.8998}
\newcommand\rmseNFprobeKurucz{0.94}
\newcommand\rmseNFprobeLimMAP{0.9227}

\author{
  \and
  Steffen Rendle\thanks{Google Research, Mountain View, USA}\\
  \texttt{srendle@google.com}
  \and
  Li Zhang\footnotemark[1]\\
  \texttt{liqzhang@google.com}
  \and
  Yehuda Koren\thanks{Google, Haifa, Israel}\\
  \texttt{yehuda@google.com}
}
\date{\vspace{-6.0ex}}

\begin{document}

\maketitle
\setcounter{footnote}{-1}

\begin{abstract}
Numerical evaluations with comparisons to baselines play a central role when judging research in recommender systems.
In this paper, we show that running baselines properly is difficult.
We demonstrate this issue on two extensively studied datasets.
First, we show that results for baselines that have been used in numerous publications over the past five years for the Movielens 10M benchmark are suboptimal.
With a careful setup of a vanilla matrix factorization baseline, we are not only able to improve upon the reported results for this baseline but even outperform the reported results of any newly proposed method.
Secondly, we recap the tremendous effort that was required by the community to obtain high quality results for simple methods on the Netflix Prize.
Our results indicate that empirical findings in research papers are questionable unless they were obtained on standardized benchmarks where baselines have been tuned extensively by the research community.
\end{abstract}

\section{Introduction}

In the field of recommendation systems, numerical evaluations play a central role for judging research.
Newly published methods are expected to be compared to baselines, i.e., well-known approaches, in order to measure the improvements over prior work.
The best practices require reproducible experiments on several datasets with a clearly described evaluation protocol, baselines tuned by hyperparameter search, and testing for statistical significance of the result.
Findings from such experiments are considered reliable.
In this work, we question this practice and show that running baselines properly is difficult.

We highlight this issue on the extensively studied Movielens 10M (ML10M) benchmark~\cite{harper:2015}.
Over the past five years, numerous new recommendation algorithms have been published in prestigious conferences such as ICML~\cite{lee:icml13,li:icml16,zheng:icml16}, NeurIPS~\cite{li:nips17}, WWW~\cite{sedhain:www15,li:www18}, SIGIR~\cite{chen:sigir15}, or AAAI~\cite{li:aaai17,chen:aaai17} reporting large improvements over baseline methods.
However, we show that with a careful setup of a vanilla matrix factorization baseline, we are not only able to outperform the reported results for this baseline but even the reported results of any newly proposed method.
Our findings question the empirical conclusions drawn from five years of work on this benchmark.
This is worrisome because the ML10M benchmark follows the best practices of reliable experiments.
Even more, if results on a well studied benchmark such as ML10M are misleading, typical one-off evaluations are more prone to producing unreliable results.
Our explanation for the failure of providing reliable results is that the difficulty of running baselines is widely ignored in our community.
This difficulty of properly running machine learning methods could be observed on the Netflix Prize~\cite{bennett:kddcup07} as well.
Section~\ref{sec:netflix} recaps the tremendous community effort that was required for achieving high quality results for vanilla matrix factorization.
Reported results for this simple method varied substantially but eventually the community arrived at well-calibrated numbers.
The Netflix Prize also highlights the benefits of rigorous experiments: the findings stand the test of time and, as this paper shows, the best performing methods of the Netflix Prize also work best on ML10M.

Recognizing the difficulty of running baselines has implications both for conducting experiments and for drawing conclusions from them.
The common practice of one-off evaluations where authors run several baselines on a few datasets is prone to suboptimal results and conclusions should be taken with care.
Instead, high confidence experiments require standardized benchmarks where baselines have been tuned extensively by the community.
Finally, while our work focuses exclusively on evaluations, we want to emphasize that empirical comparisons using fixed metrics are not the only way to judge work.
For example, despite comparing to sub-optimal baselines on the ML10M benchmark, recent research has produced many useful techniques, such as local low rank structures~\cite{lee:icml13}, mixtures of matrix approximation~\cite{li:nips17}, and autoencoders~\cite{sedhain:www15}.

\section{Observations}
\label{sec:experiments}

In this section, we first examine the commonly used Movielens 10M benchmark for rating prediction algorithms~\cite{lee:icml13,sedhain:www15,chen:sigir15,zheng:icml16,strub:arxiv16,chen:ijcai16,li:icml16,chen:aaai17,li:nips17,li:www18}.
We show that by carefully setting up well known methods, we can largely outperform previously reported results.
The surprising results include (1)~Bayesian MF~\cite{salakhutdinov:icml08,freudenthaler:nipsws11} which was reported by previous authors to be one of the poorest performing methods, can outperform any result reported so far on this benchmark including all of the recently proposed algorithms.
(2)~Well known enhancements, proposed a decade ago for the Netflix Prize, such as SVD++~\cite{koren:kdd08} or timeSVD++~\cite{koren:kdd09}, can further improve the quality substantially.

Secondly, we compare these findings to the well documented evolution of results on the Netflix Prize.
Also on the Netflix Prize, we observe that setting up methods is challenging.
For example, results reported for matrix factorization vary considerably over different publications~\cite{funksvd,kurucz:kddcup07,lim:kddcup07,takacs:kddcup07,paterek:kddcup07,salakhutdinov:icml08,pilaszy:rs10,bell:progressprize07,zhou:aaim08}.
However, the Netflix Prize encouraged tweaking and reporting better runs of the same method, so over the long run, the results were well calibrated.

\subsection{Movielens}
\label{sec:movielens}

\begin{figure}[t]
    \centering
    \includegraphics[height=2.9in]{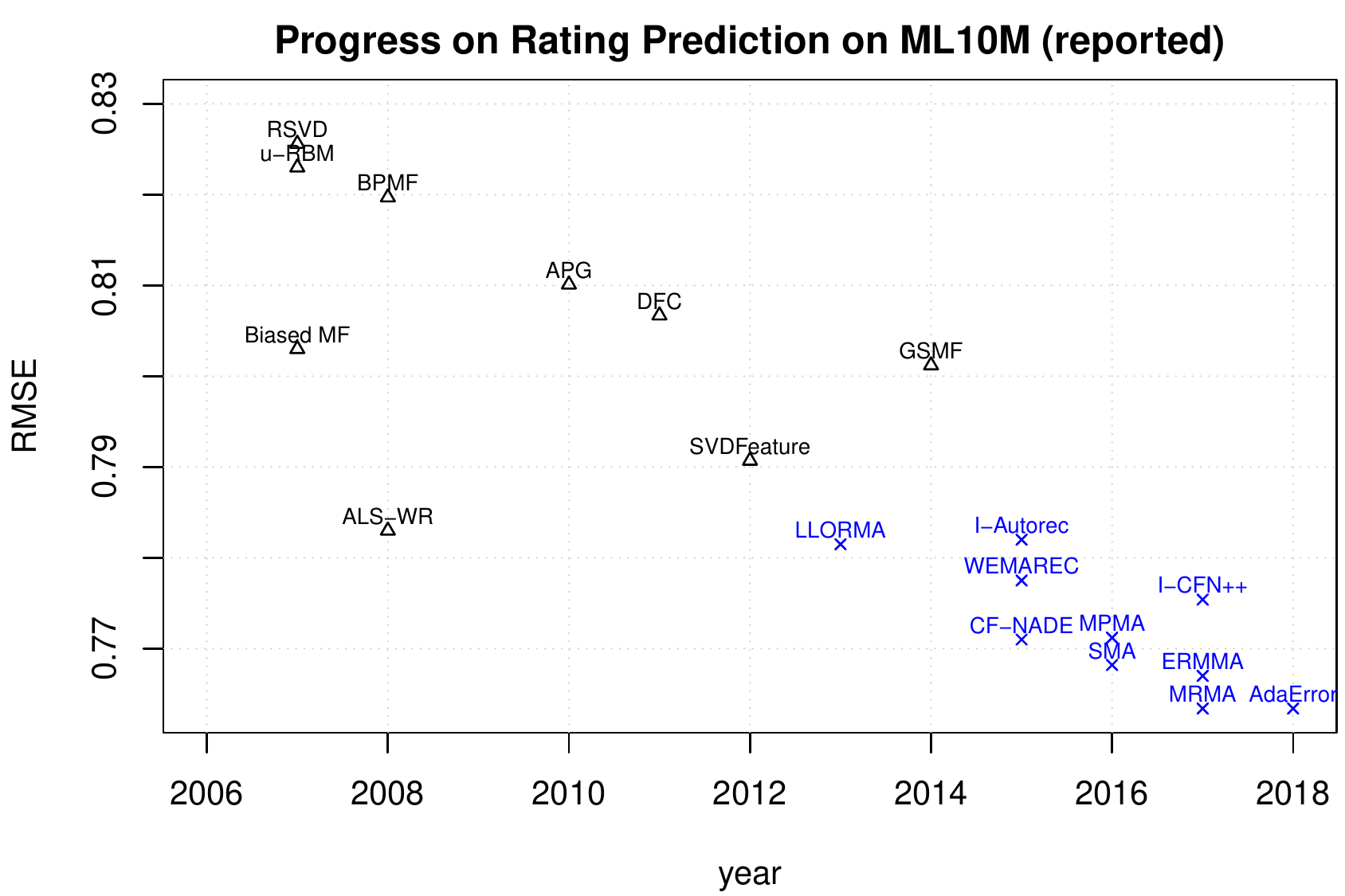}
    \caption{Progress on rating prediction measured on the Movielens 10M benchmark.
    Results marked as blue crosses were reported by the corresponding inventors.
    Results marked as black triangles were run as baselines by authors of newly invented methods.
    Results are from \cite{lee:icml13,sedhain:www15,chen:sigir15,zheng:icml16,strub:arxiv16,chen:ijcai16,li:icml16,chen:aaai17,li:nips17,li:www18}. See Table~\ref{tbl:ml10m} for details.}
    \label{fig:ml10m_progress}
\end{figure}

Measuring Root Mean Square Error (RMSE) on a global random 90:10 split of Movielens 10M is a common benchmark for evaluating  rating prediction methods~\cite{lee:icml13,sedhain:www15,chen:sigir15,zheng:icml16,strub:arxiv16,chen:ijcai16,li:icml16,chen:aaai17,li:nips17}.
Figure~\ref{fig:ml10m_progress} shows the progress reported over the past 5 years on this benchmark.
All newly proposed methods clearly outperform the earlier baselines such as matrix factorization or Boltzman machines~\cite{salakhutdinov:icml07} (RBM).
Both SGD (RSVD, Biased MF) and Bayesian versions (BPMF) of matrix factorization have been found to perform poorly.
The figure indicates a steady progress, by improving the state-of-the art in rating prediction considerably.
Many results include also standard deviations to show that the results are statistically significant, e.g.,~\cite{strub:arxiv16,li:nips17}.

\subsubsection{A Closer Look at Baselines}

The reported results for Biased MF, RSVD, ALS-WR, and BPMF indicate some issues.
\begin{enumerate}
    \item Biased MF~\cite{koren:ieee09} and RSVD~\cite{paterek:kddcup07} are essentially the same method: L2 regularized matrix factorization learned with SGD.
Qualitative differences should only stem from different setups such as hyperparameters, training data ordering, or implementations.
    \item ALS-WR~\cite{zhou:aaim08} and Biased MF/ RSVD are identical models learned by different algorithms.
When both are tuned well, they have shown very similar results on the Netflix Prize (see Section~\ref{sec:netflix}).
    \item BPMF~\cite{salakhutdinov:icml08} shares the model with RSVD/ALS-WR but is learned with a Gibbs sampler.
On the Netflix Prize it has shown the best performance for learning matrix factorization models \cite{salakhutdinov:icml08,rendle:vldb13} compared to other learning methods (SGD, ALS, VB).
It is surprising that it shows much worse quality than Biased MF and ALS-WR on Movielens 10M.
\end{enumerate}

\subsubsection{Rerunning Baselines}

\begin{figure}[t]
    \centering
    \includegraphics[height=2.9in]{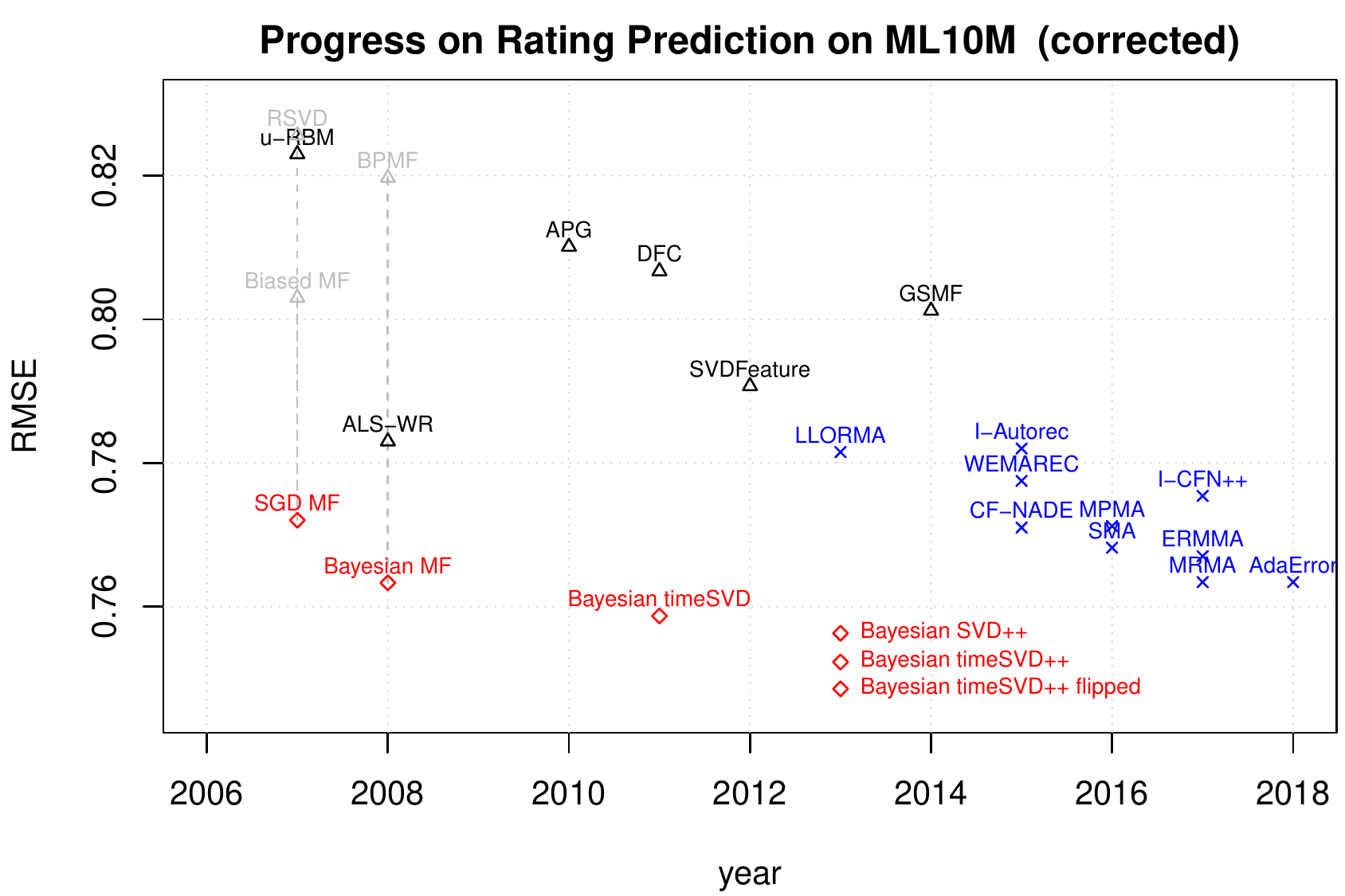}
    \caption{Rerunning baselines for Bayesian MF, and adding popular methods such as Bayesian versions of SVD++, timeSVD++, timeSVD\protect\footnotemark.
    Bayesian MF (=BPMF) and SGD MF (=RSVD, Biased MF) can achieve much better results than previously reported.
    With a proper setup, well known methods can outperform any recently proposed method on this benchmark.}
    \label{fig:ml10m_progress_corrected}
\end{figure}

We reran the baselines and a different picture emerged (see Figure~\ref{fig:ml10m_progress_corrected} and Table~\ref{tbl:ml10m}).
More details about the experiments can be found in the Appendix.

\footnotetext{SVD++ was introduced in 2008~\cite{koren:kdd08}, timeSVD++ in 2009~\cite{koren:kdd09}.
We use here a Bayesian FM solver which was introduced in 2011~\cite{freudenthaler:nipsws11}.
The Bayesian solver to handle the implicit case efficiently (i.e., the "++") was published in 2013~\cite{rendle:vldb13}.}

\paragraph{Matrix Factorization}

First, we ran a matrix factorization model learned by an SGD algorithm (similar to RSVD and Biased MF).
SGD-MF achieved an RMSE of {\rmsefmui} for a 512-dimensional embedding and an RMSE of {\rmsefmuimed} for 64 dimensions.
This is considerably better than the reported values for RSVD ({\rmsersvd}) and Biased MF ({\rmsebiasmf}), and outperforms even several of the newer methods such as LLORMA ({\rmsellorma}), Autorec ({\rmseiautorec}), Wemarec ({\rmsewemarec}), or I-CFN++ ({\rmseicfnpp}).

Next, we trained a Bayesian matrix factorization model using Gibbs sampling (similar to BPMF).
Bayesian MF achieved an RMSE score of {\rmsebfmui} for a 512 dimensional embedding.
This is not only much better than the previously reported~\cite{li:icml16,strub:arxiv16} number for BPMF ({\rmsebpmf}) but it outperforms even the best method (MRMA {\rmsemrma}) ever reported on ML10M.

\paragraph{Stronger Baselines}

One of the lessons of the Netflix Prize was that modeling implicit activity was highly predictive and outperformed vanilla matrix factorization.
SVD++\cite{koren:kdd08}, the asymmetric model (NSVD1)~\cite{paterek:kddcup07} and to some extent RBMs~\cite{salakhutdinov:icml07} are examples of models harnessing implicit feedback.
Another important aspect was capturing temporal effects~\cite{koren:kdd09}.

First, we added a time variable to the Bayesian matrix factorization model, and achieved an RMSE of \rmsebfmuit.
Second, we trained an implicit model by adding a bag-of-words predictor variable that includes all the videos a user watched.
This model is equivalent to SVD++~\cite{koren:kdd08,rendle:tist12}.
It further improved over Bayesian MF, and achieves an RMSE of \rmsebfmuiui.
Next, we trained a joint model with time and the implicit usage information, similar to the timeSVD++ model~\cite{koren:kdd09}.
This model achieved an RMSE of \rmsebfmuitui.
Finally, we added a flipped version of the implicit usage in timeSVD++: a bag of word variable indicating the other users who have watched a video.
This model dropped the RMSE to \rmsebfmuituiiu.

\paragraph{Summary}
By carefully setting up baselines, we could outperform any result even with a simple Bayesian matrix factorization -- a method that was previously reported to perform poorly on ML10M.
Applying modeling techniques known for almost a decade, we were able to achieve substantial improvements -- in absolute terms, we improved over the previously best reported result, MRMA, from 2017 by 0.0144, a similar margin as several years of improvements reported on this dataset\footnote{The difference between LLORMA in 2013 to MRMA in 2017 is 0.0186.}.
Our results question conclusions drawn from previous experimental results on ML10M.
Instead of improving over the baselines by a large margin, all recently proposed methods \emph{underperform} well-known baselines substantially.

\pagebreak[4]

\begin{table}[h]
\centering
\begin{tabular}{lSl}
\toprule
Method & {RMSE} & Comment \\
\midrule
RSVD~\cite{paterek:kddcup07} & \rmsersvd & result from \cite{li:icml16}\\
U-RBM~\cite{salakhutdinov:icml07} & \rmseurbm & result from \cite{sedhain:www15} \\
BPMF~\cite{salakhutdinov:icml08} & \rmsebpmf & result from \cite{li:icml16}\\
APG~\cite{toh:pjo10} & \rmseapg & result from \cite{li:icml16}\\
DFC~\cite{mackey:nips11} & \rmsedfc & result from \cite{li:icml16} \\
Biased MF~\cite{koren:ieee09} & \rmsebiasmf & result from \cite{sedhain:www15}\\
GSMF~\cite{yuan:aaai14} & \rmsegsmf & result from \cite{li:icml16}\\
SVDFeature~\cite{chen:jmlr12} & \rmsesvdfeature & result from \cite{strub:arxiv16} \\ 
ALS-WR~\cite{zhou:aaim08} & \rmsealswr & result from \cite{strub:arxiv16} \\
\midrule
I-AUTOREC~\cite{sedhain:www15} & \rmseiautorec &  result from \cite{sedhain:www15} \\
LLORMA~\cite{lee:icml13} & \rmsellorma & result from \cite{lee:icml13} \\
WEMAREC~\cite{chen:sigir15} & \rmsewemarec & result from \cite{chen:sigir15}\\
I-CFN++~\cite{strub:arxiv16} & \rmseicfnpp & result from \cite{strub:arxiv16} \\ 
MPMA~\cite{chen:ijcai16} & \rmsempma & result from \cite{chen:ijcai16} \\
CF-NADE 2 layers~ \cite{zheng:icml16} & \rmsenade & result from \cite{zheng:icml16} \\
SMA~\cite{li:icml16} & \rmsesma & result from \cite{li:icml16}\\
GLOMA~\cite{chen:aaai17} & \rmsegloma & result from \cite{chen:aaai17} \\
ERMMA~\cite{li:aaai17} & \rmseermma & result from \cite{li:aaai17} \\
AdaError~\cite{li:www18} & \rmseadaerror & result from \cite{li:www18} \\
MRMA~\cite{li:nips17} & \rmsemrma & result from \cite{li:nips17} \\
\midrule
SGD MF~\cite{paterek:kddcup07,koren:ieee09} & \rmsefmui & same method as RSVD, Biased MF \\
Bayesian MF~\cite{salakhutdinov:icml08,freudenthaler:nipsws11} & \rmsebfmui & same method as BPMF\\
Bayesian timeSVD~\cite{koren:kdd09,freudenthaler:nipsws11,rendle:vldb13} & \rmsebfmuit & MF with a time variable \\
Bayesian SVD++~\cite{koren:kdd08,rendle:vldb13} & \rmsebfmuiui & similar to \cite{koren:kdd08} learned with MCMC\\
Bayesian timeSVD++~\cite{koren:kdd09,rendle:vldb13} & \rmsebfmuitui & similar to \cite{koren:kdd09} learned with MCMC \\
Bayesian timeSVD++ flipped~\cite{rendle:vldb13} & \rmsebfmuituiiu & adding implicit item information \\
\bottomrule
\end{tabular}
    \caption{Movielens 10M results: first group are baselines. Second group are newly proposed methods. Third group are baseline results that we reran. See Appendix for details of our results.
    \label{tbl:ml10m}}
\end{table}

\pagebreak[4]

\subsection{Netflix Prize}
\label{sec:netflix}

The Netflix Prize~\cite{bennett:kddcup07} also indicates that running methods properly is hard.
We are highlighting this issue by revisiting the large community effort it took to get well calibrated results for the vanilla matrix factorization model.

\subsubsection{Background}

The Netflix Prize was awarded to the first team that decreases by 10\% the RMSE of Netflix's own recommender system, Cinematch, with an RMSE score of $0.9514$~\cite{bennett:kddcup07}.
It took the community about three years and hundreds of ensembled models to beat this benchmark.
Given that the overall relative improvement for winning the prize was $0.095$, a difference of $0.01$ in RMSE scores is considered large -- e.g., it took one year to lower the RMSE from $0.8712$ (\emph{progress prize 2007}) to $0.8616$ (\emph{progress prize 2008}) and the \emph{Grand prize} was awarded in 2009 for an RMSE of $0.8554$.

The Netflix Prize dataset is split into three sets: a training set, the \emph{probe} set for validation and the \emph{qualifying} set for testing.
The ratings of the qualifying set were withheld during the competition.
Participants of the Netflix Prize could submit their predictions on the qualifying set to the organizer and the RMSE score on half of this set, the \emph{quiz} set, was reported on the \emph{public} leaderboard.
The RMSE on the remaining half, the \emph{test} set, was private to the organizer and used to determine the winner.
Scientific papers usually report either the probe RMSE or the quiz RMSE\footnote{The quiz set is easier to predict than the probe set (typically, RMSE is about $0.007$ lower), because for training the \emph{quiz model}, the probe set can be added to enrich the training set.}.

The data was split based on user and time between the training, probe and qualifying set.
In particular, for every user, the last six ratings were withheld from training, three of them were put into the probe set and three into the qualifying set.
This split technique is more challenging than global random splitting\footnote{Unfortunately, in recent work on rating prediction it is common to evaluate on a global random 90:10 split of the Netflix Prize data.
Results of global splitting are not comparable to numbers reported on the Netflix Prize split.
This is unfortunate because the original Netflix split has been extensively studied and has very well calibrated results.} for two reasons.
(1)~Users with few ratings have the same number of evaluation data points as frequent users.
Compared to a global random split, ratings by users with little training data, i.e., harder test cases, are overrepresented in the test dataset.
(2)~Withholding by time makes this a forecasting problem where test ratings have to be extrapolated whereas a global random split allows simpler interpolation.

\subsubsection{Matrix Factorization}

From the beginning of the competition, matrix factorization algorithms were identified as promising methods.
Very early results used traditional SVD solvers with sophisticated imputation and reported results close to a probe RMSE of \rmseNFprobeKurucz~\cite{kurucz:kddcup07}.
A breakthrough was FunkSVD~\cite{funksvd}, a sparse matrix factorization algorithm that ignored the missing values.
It was learned by iterative SGD with L2 regularization and achieved a probe RMSE of $0.93$.
This encouraged more researchers to experiment with matrix factorization models and in the KDDCup 2007 workshop, participants reported improved results of \rmseNFprobeLimMAP~(probe)~\cite{lim:kddcup07}, \rmseNFprobeGravity~(probe)~\cite{paterek:kddcup07}, and \rmseNFtestpatarek~(quiz)~\cite{paterek:kddcup07}.
Participants continued to improve the results for the basic matrix factorization models and reported scores as low as \rmseNFtestZhou~\cite{zhou:aaim08} for ALS and \rmseNFtestYehuda~\cite{koren:grandprize} for SGD methods.

\begin{table}[t]
    \centering
    \begin{tabular}{lll} \\
    \hline
    Team / Publication & {Probe RMSE} & {Quiz RMSE}\\
    \hline
    Kurucz et al.~\cite{kurucz:kddcup07} & \rmseNFprobeKurucz & -  \\
    Simon Funk \cite{funksvd} & \rmseNFprobefunksvd &  - \\
    Lim and Teh~\cite{lim:kddcup07}          & \rmseNFprobeLimMAP & - \\
    Gravity~\cite{takacs:kddcup07}          & \rmseNFprobeGravity & - \\
    Paterek~\cite{paterek:kddcup07} & - & \rmseNFtestpatarek\\
    Pragmatic Theory~\cite{pt:grandprize} & \rmseNFprobePT & \rmseNFtestPT \\
    Big Chaos~\cite{bigchaos:grandprize} & - & \rmseNFtestbgSGD \\
    Pilaszy et al.~\cite{pilaszy:rs10}     & - & \rmseNFtestPilaszy \\
    BellKor~\cite{bell:progressprize07} & & \rmseNFtestYehuda \\
    Zhou et al.~\cite{zhou:aaim08} & - & \rmseNFtestZhou \\
    \bottomrule
    \end{tabular}
    \caption{Netflix Prize: Results for vanilla matrix factorization models using ALS and SGD optimization methods.}
    \label{tbl:netflix_mf}
\end{table}

Table~\ref{tbl:netflix_mf} summarizes some of the key results for vanilla matrix factorization including the results reported by top competitors and the winners.
These results highlight that achieving good results even for a presumably simple method like matrix factorization is non trivial and takes large effort.

\subsubsection{Refinements and Winning Algorithms}

Our previous discussion was restricted to vanilla matrix factorization models.
After the community converged to well calibrated results for matrix factorization, the focus shifted to more complex models taking into account additional information such as implicit feedback (e.g., SVD++~\cite{koren:kdd08}) and time (e.g., timeSVD++~\cite{koren:kdd09}).
The most sophisticated timeSVD++ models achieved RMSEs as low as 0.8762~\cite{koren:grandprize}.

All the top performing teams also relied heavily on ensembling as many diverse models as possible including sophisticated nearest neighbor models~\cite{koren:ieee09}, or Restricted Boltzmann Machines~\cite{salakhutdinov:icml07}.
The final models that won the Netflix Prize were ensembles of several teams each with dozens of models~\cite{koren:grandprize}.

\subsection{Discussion}

Compared to recent evaluations on ML10M, the Netflix Prize encouraged rerunning methods and reporting improvements on identical methods (see Table~\ref{tbl:netflix_mf}).
This ensured that the community converged towards understanding which methods work well.
In contrast to this, for ML10M there is no encouragement to rerun results for simple baselines, or even to outperform complex new methods.
One explanation is that for the Netflix Prize, participants get rewarded by getting a low RMSE -- no matter how it was achieved.
In terms of publications, which motivates current work on rating prediction, achieving good results with old approaches is usually not seen as a scientific contribution worth publishing.

However, the ultimate goal of empirical comparisons is to better understand the trade-offs between alternative methods and to draw insights about which patterns lead to successful methods.
Our experiments have shown that previous empirical results on ML10M fail to deliver these insights.
Methods that were reported to perform poorly actually perform very well.
In contrast to this, our experiments on ML10M show that all the patterns learned on the Netflix Prize hold also on ML10M, and the best Netflix Prize methods perform also best on ML10M.
In this sense, the Netflix Prize was successful and ML10M benchmark was not (so far).

Like previous baselines were not properly tuned on ML10M, it is possible that the recently proposed methods could also improve their results with a more careful tuning.
This would not be a contradiction to our observation but be further evidence that running experiments is hard and needs large effort of experimentation and tuning to achieve reliable results.

Finally, we want to stress that this is not a unique issue of ML10M.
Quite the opposite, most work in recommendation systems is not even evaluated on a standardized benchmark such as ML10M.
Results obtained on one-off evaluations are more prone to questionable experimental findings than ML10M.

\section{Insufficient Indicators for Experimental Reliability}

We shortly discuss commonly used indicators that have been used to judge the reliability of experimental results, such as statistical significance, reproducibility or hyperparameter search.
While all of them are necessary, we argue that they are not sufficient to ensure reliable results.
Our results in Section~\ref{sec:experiments} suggest that they are less important than proper set ups.

\subsection{Statistical Significance}

Most of the results for ML10M are reported with a standard deviation (e.g, \cite{strub:arxiv16,li:nips17}).
The reported standard deviation is usually low and the difference of the reported results are \emph{statistically significant}.
Even for the reported BPMF results in \cite{li:nips17}, the standard deviation is low (0.0004).
Based on our study (Section \ref{sec:movielens}), statistically significant results should not be misinterpreted as a ``proof'' that method A is better than B.
While this sounds like a contradiction, statistical significance does not measure how well a method is set up.
It measures the variance within one setup.

Statistical significance and standard deviations should only be considered after we have evidence that the method is used well.
We argue that setting the method up properly is a much larger source for errors.
In this sense, statistical significance is of little help and often provides a false confidence in experimental results.

\subsection{Reproducibility}

The ability to rerun experiments and to achieve the same numbers as reported in previous work is commonly referred to as \emph{reproducibility}.
Often, implementations and hyperparameter setups are shared by authors to allow reproducing results.
While reproducibility is important, it does not solve the issue we point out in this work.
Rerunning the code of authors can reproduce the results but it is not evidence of a proper setup.
In the example of the ML10M dataset, the dataset is public, the experimental protocol is documented and simple, and there exist plenty implementations of the baselines -- even authors commonly make their new methods public.
The same holds for the Netflix prize, or most machine learning competitions.
Despite the easy reproducibility, the conclusions from experimental results can be questionable (see Section~\ref{sec:movielens}).

\subsection{Tuned Hyperparameters}

One of our central arguments is that it is not easy to run a machine learning method properly.
In most research papers, it is common practice to search over the hyperparameter space (e.g., learning rates, embedding dimension, regularization) and report the results for the ``best'' setting.
However, Section~\ref{sec:experiments} indicates that this still does not solve the problem, and reported results can vary substantially from a proper setup.
We speculate that hyperparameter search spaces are often incomplete and do not replace experience with a method.
For example, interpreting and acting on the results of different hyperparameter settings is non-trivial, e.g., should the boundary be extended, or refined? What is the right search grid?
Can we search hyperparameters on a small model and transfer the results to a larger one?
All these questions make it hard for setting up an unknown "black-box".

A second source are knobs that are not even considered during hyperparameter search.
For example, a method might require to recenter the data before running it, or that the training data is shuffled, or to stop training early, or to use a certain initialization.
Such knobs might be trivial and not even worth reporting for someone with experience in a method, but will make it almost impossible for others to set comparisons up properly.
This becomes a problem when the method is run by a non-expert on a different dataset or experimental setup.

\section{Improving Experimental Quality}

Based on our findings, reliable experiments are hard to achieve by authors of a single paper but require a community effort.
We see two key requirements for this: (1)~standardized benchmarks; and (2)~incentives to run and improve baseline results.

\subsection{Standardized Benchmarks}

While today's best practices encourage authors of a paper to run as many baselines as possible, our findings indicate that this should be discouraged because it is prone to producing unreliable results.
If running baselines from scratch is discouraged, the only way to get points of comparisons to other methods are standardized benchmarks, i.e., datasets with well-defined train--test splits and evaluation protocols.
ML10M with 10 fold CV or the Netflix Prize split, both measured on RMSE, are examples of well defined benchmarks for comparing rating prediction algorithms.
However, recommender tasks are diverse, e.g. item recommendation vs. rating prediction, cold-start vs. active users, forecasting, explanation, etc. and most of them miss standardized benchmarks.
While it is important to explore new tasks, over time it is crucial for the community to converge to standardized benchmarks for reoccurring problems.
As we have argued in this paper, empirical findings on non-standardized benchmarks are likely less reliable.

A common concern about benchmarks is that methods "overfit" to a particular dataset, leading to false discoveries.
However, this is less of a problem for the scale of the data typically used in machine learning tasks.
For example, one of the most heavily researched datasets, the Netflix Prize, has shown very little signs of overfitting after more than 10 years of study.
Both the public leaderboard\footnote{\url{https://www.netflixprize.com/leaderboard_quiz.html}} and the private (hidden) leaderboard\footnote{\url{https://www.netflixprize.com/leaderboard.html}} show only minor differences in ordering and the same relative improvements.
Also, our results on the ML10M dataset (see Section~\ref{sec:movielens}) emphasize that the lessons learned on the Netflix Prize still hold after a decade and the patterns and methods that worked the best for Netflix Prize are also those best performing on ML10M.
While signs of overfitting might show up in the long run, the benefits from well calibrated results outweight issues that improper baselines might cause.

\subsection{Incentives for Running Baselines}

ML10M and the Netflix Prize are both standardized benchmarks.
However, one of them produced well calibrated results, while the other one had misleading baselines for many years (see Table~\ref{tbl:ml10m}).
Our explanation of this phenomena is that there is no encouragement to keep on improving baselines for ML10M.
\emph{Novelty} is a key criterion to judge research contributions.
Achieving good results with a well known method gets little reward, so researchers do not spend much effort on it -- and even if good results would be achieved, it is hard to publish them.
In contrast to this, the Netflix Prize encouraged spending time on experimenting with existing methods.
This was the most certain way of getting good results, and a chance to improve on the contest leaderboard.
Real life systems often also incentivize bettering known, well established methods rather than inventing new ones.
We think it is crucial to find incentives for tuning well-known methods on benchmarks.
As we have shown, this is a non-trivial task which needs expertise and time.
Without well calibrated results, conclusions drawn from experiments are questionable.

Besides evaluations motivated by scientific publications, machine learning competitions, e.g., on platforms such as Kaggle\footnote{\url{http://www.kaggle.com/}} or organized by conferences such as the annual KDDCup, can serve as standardized benchmarks with well-calibrated results.

\section{Conclusion}

In this paper, we have shown that results for baselines that have been used in numerous publications over the past five years for the ML10M benchmark are suboptimal.
With a careful setup of a vanilla matrix factorization baseline, we were not only able to outperform the reported results for the baselines but even the reported results of any newly proposed method.
Other well-known models such as SVD++ provide an even higher gain.
These results are surprising as the papers follow the best practices in our community to ensure reliable results:
they conduct a reasonable hyperparameter search, report statistical significance and allow reproducibility.
This indicates that running baseline methods properly is difficult.
As recommender systems evaluation relies heavily on empirical results, the shortcomings discussed in this work highlight a major issue in our ability to judge work.
Our findings question the common practice in recommender systems research papers of running baseline models and experimenting on multiple datasets.
Even when following the best practices as outlined above, results can be unreliable.
Our work indicates that trustworthy baselines require standardized benchmarks and considerable tuning effort by the community.

\newpage\section*{Appendix}

In this section, we describe our experiments and setup in more detail.
We experiment on the Movielens 10M dataset\footnote{\url{http://grouplens.org/datasets/movielens/10m/}} with a 10 fold cross validation protocol. That is, 10 random splits each with 90\% training data and 10\% test data, where the 10 test splits do not overlap.
Our test protocol allows to compare our results to previous publications on the Movielens 10M dataset with a random 90:10 split~\cite{lee:icml13,sedhain:www15,li:icml16,strub:arxiv16,chen:ijcai16,zheng:icml16,chen:aaai17,li:nips17}.
Our experiments focus mainly on Bayesian models learned by Gibbs sampling, a Markov Chain Monte Carlo method, because they have fewer critical hyperparameters than SGD.
However, we also run SGD matrix factorization to be able to compare to existing numbers~\cite{sedhain:www15,li:icml16} reported for this method.

\subsection{Factorization Models}
\label{sec:fm_features}

\begin{table}[t]
\begin{center}
\begin{tabular}{llp{0.5\textwidth}}
\toprule
Name &  Features &  Comment \\
\midrule
Matrix Factorization & u, i &  Equivalent to biased matrix factorization~\cite{koren:ieee09} and RSVD~\cite{paterek:kddcup07} \\
SVD++ & u, i, iu  &  Similar to \cite{koren:kdd08} \\
timeSVD & u, i, t &  Similar to \cite{koren:kdd09} \\
timeSVD++ & u, i, t, iu & Similar to \cite{koren:kdd09} \\
timeSVD++ flipped & u, i, t, iu, ii  &\\
\bottomrule
\end{tabular}
\end{center}
   \caption{Models from our ML10M experiment and the corresponding features that we used in a Factorization Machine. See Section~\ref{sec:fm_features} for an explanation of the features.}
    \label{tbl:models}
\end{table}

We used the factorization machine library, libFM\footnote{\url{https://github.com/srendle/libfm}}~\cite{rendle:tist12}, for all experiments.
We consider five features which have been used successfully on the Netflix Prize:
\begin{enumerate}
    \item User id (u): a categorical variable indicating the user
    \item Item id (i): a categorical variable indicating the movie
    \item Time (t): the day of the rating, which is treated as a categorical variable with one category per day
    \item Implicit user information (iu): a bag of words variable that is the set of all movie ids a user has ever watched\footnote{We used the same protocol as in the Netflix prize and included \emph{which} movies the user rated from both the training and test set -- for sure, we did not include the rating of the test set. We also experimented with a stricter definition of implicit information and removed information about which movies a user rated in the test set. As expected, this resulted in slightly worse RMSE numbers. However, implicit information is still very useful in the stricter setting.}
    \item Implicit item information (ii): a bag of words variable that is the set of all user ids that have ever watched a movie
\end{enumerate}
Table~\ref{tbl:models} lists the combination of features that we used.

\subsection{Bayesian Learning}

\begin{figure}[t]
    \centering
    \includegraphics[height=3in]{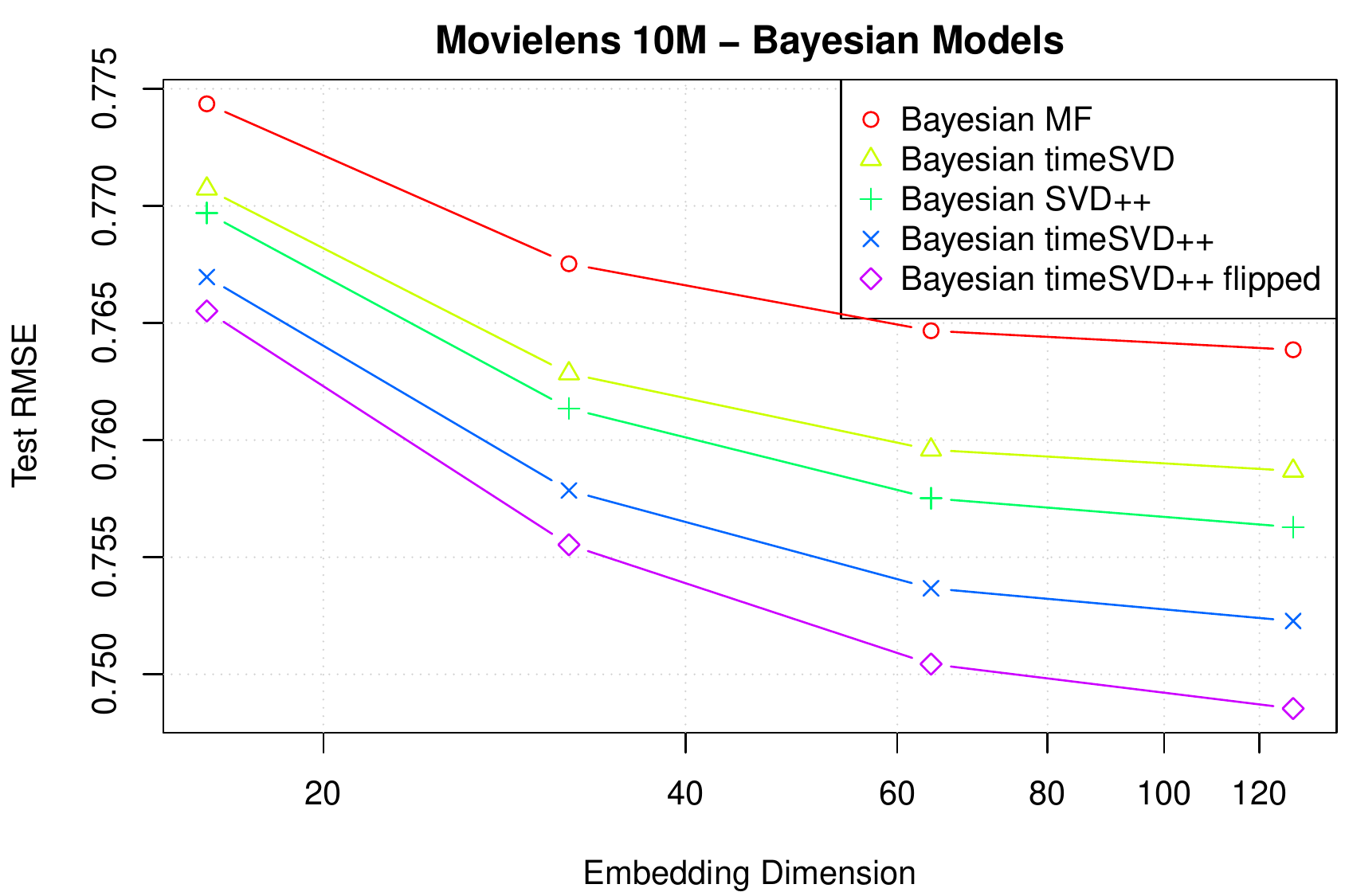}
    \caption{Quality of the Bayesian models with an increasing embedding dimension (with $512$ sampling steps). Larger dimensions show better quality.}
    \label{fig:ml10m_quality_vs_dim}
\end{figure}

\begin{figure}[t]
    \centering
    \includegraphics[height=3in]{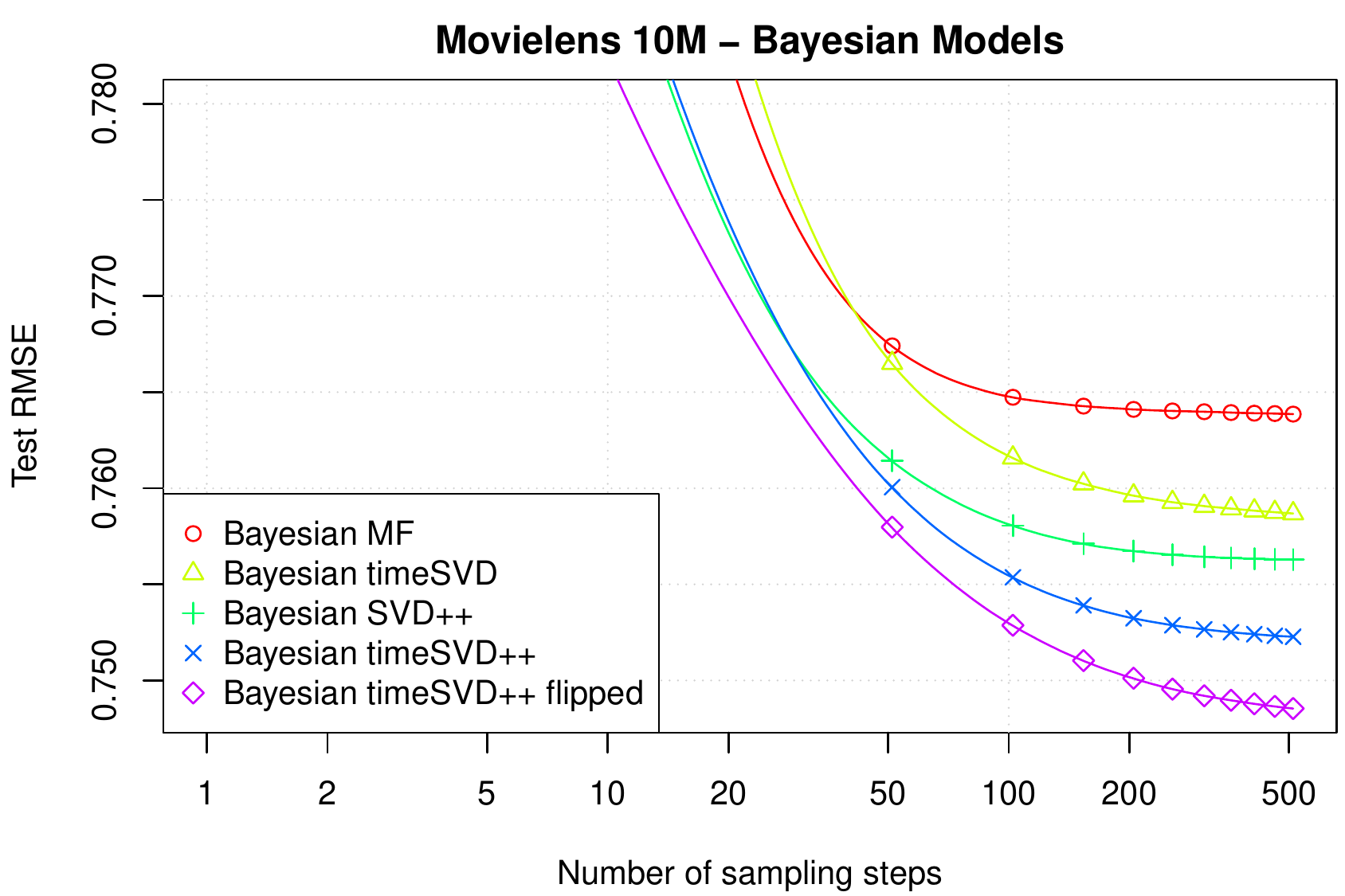}
    \caption{Quality of the Bayesian models with an increasing number of sampling steps (with $128$ dimensional embeddings). More sampling steps show better quality.}
    \label{fig:ml10m_quality_vs_steps}
\end{figure}

Setting up a Bayesian model is very simple.
There are three critical settings: (a)~number of sampling steps, (b)~dimension of the embedding, and (c)~initialization of model parameters.
In our experience, the more sampling steps and the higher the dimension the better the quality.
We report results for up to $512$ steps (iterations), and embedding dimensions of $16$, $32$, $64$, and $128$ -- for matrix factorization, we also ran with embedding dimensions of $256$ and $512$.
For the initialization, we choose a random initialization from a Gaussian distribution with standard deviation of $0.1$.
The value $0.1$ is the default in libFM and has worked well in the Netflix prize~\cite{rendle:vldb13} and also for other Movielens splits~\cite{yamada:kdd17}.
We use the relational data representation and solver of libFM~\cite{rendle:vldb13}.

Figure~\ref{fig:ml10m_quality_vs_dim} shows the final test RMSE vs. the embedding dimension for these models.
The plot confirms that increasing the embedding dimension helps.
It also shows that features that worked well in the Netflix prize, achieve high quality on Movielens 10M.
The convergence graph, Figure~\ref{fig:ml10m_quality_vs_steps}, confirms that the prediction quality improves with more sampling steps.

\subsection{Stochastic Gradient Descent}

\begin{figure}[t]
    \centering
    \includegraphics[height=3in]{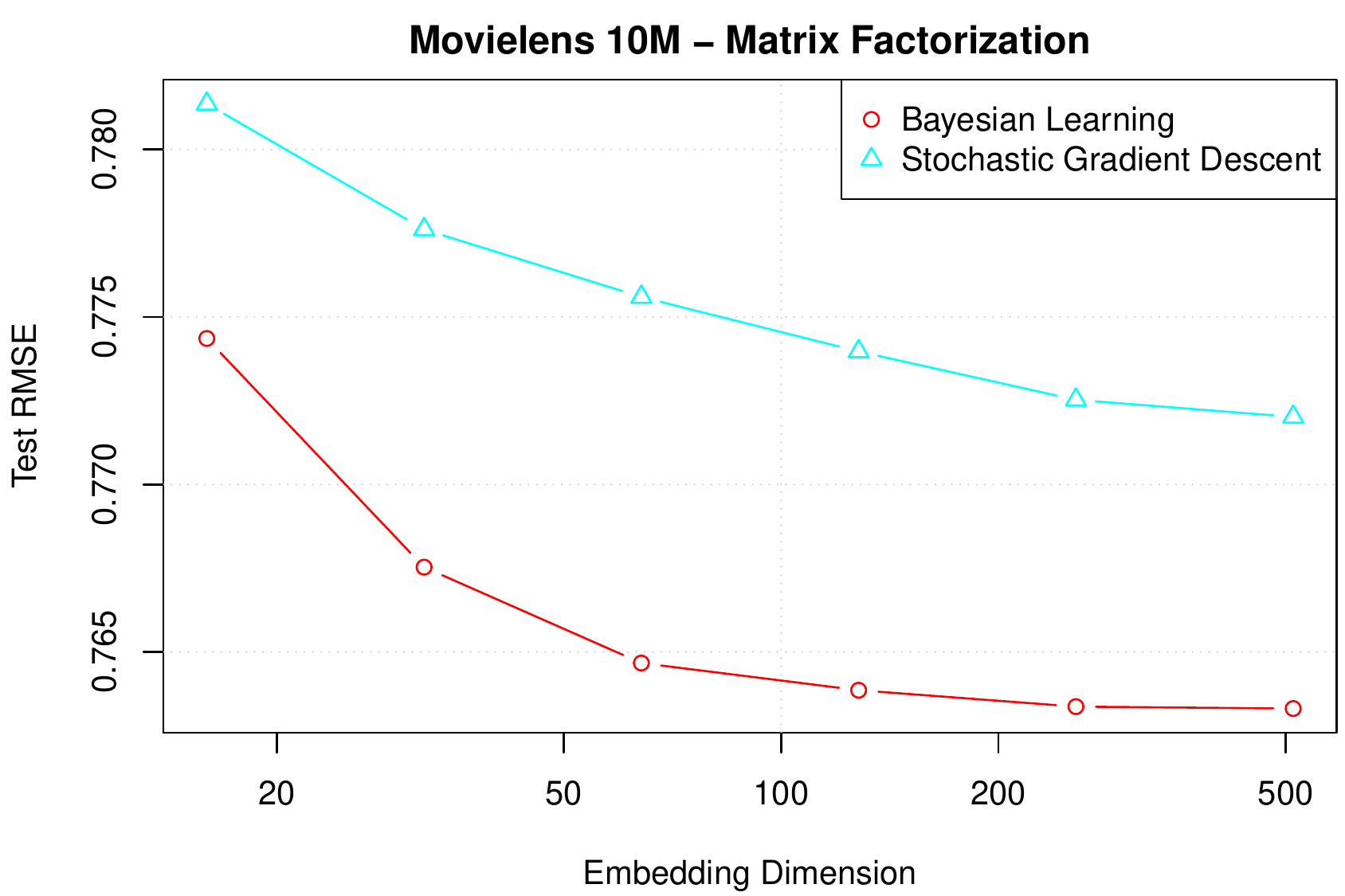}
    \caption{Comparison of Matrix Factorization learned by Gibbs Sampling (Bayesian Learning) and stochastic gradient descent (SGD) for an embedding dimension from 16 to 512.}
    \label{fig:ml10m_algorithm}
\end{figure}
For SGD, we experimented only with matrix factorization.
Compared to the Bayesian models, our SGD implementation has two additional hyperparameters: the learning rate, and regularization.
In our experience, the smaller the learning rate the better the quality, however, the number of iterations will also grow.
That means learning rate and number of iterations are not independent settings but form a runtime trade-off.
We fix the number of iterations to $128$ epochs and search for the best learning rate within this computational budget.
Also for the embedding dimension, the larger the dimension the better the quality, provided that the regularization value and number of iterations is set properly.
Larger dimensions are more costly, so we report numbers for $16$, $32$, $64$, $128$, $256$, and $512$ dimensions.
For Normal initialization, we pick $0.1$ as the standard deviation, which is the same value as for the Bayesian methods.

That leaves us with two hyperparameters to tune: (a)~learning rate, and (b)~regularization value.
We perform hyperparameter selection on the training set.
We use $5\%$ of the training set for validating hyperparameter choices and the remaining $95\%$ for training.
We set up a grid over four regularization values $\{0.02, 0.03, 0.04, 0.05\}$ and two learning rates $\{0.001, 0.003\}$.
This range of values was motivated by successful SGD hyperparameter combinations for matrix factorization on the Netflix prize~\cite{paterek:kddcup07,koren:kdd08}.
We picked $64$ dimensions for tuning the hyperparameters because it is sufficiently large to show the importance of the regularization value but small enough that the computational cost of hyperparameter search is reasonable.

The selected hyperparameters were stable among folds and for all 10 folds, the best regularization value was $0.04$ and the best learning rate $0.003$.
Finally, using the previously selected hyperparameters, we trained models for $16$, $32$, $64$, $128$, $256$, and $512$ dimensions on the full training data and evaluated on the $10\%$ test split.
Figure~\ref{fig:ml10m_algorithm} shows the final quality of matrix factorization models learned by SGD and MCMC from $16$ to $512$ dimensions.
The picture confirms that the larger the dimension, the better the quality.

It is likely that more sophisticated hyperparameter selection might lead to better SGD performance.
For example, for the Netflix prize, we observed that using individual learning rates and regularization values for biases and embeddings as well as users and items can further improve results. In addition, a decay schedule of learning rates, which decreases them at later iterations, is also known to improve accuracy.
Nevertheless, even with the above described hyperparameter search, we were able to outperform previously reported results for SGD substantially.

\end{document}